\documentclass[prl,aps,showpacs,onecolumn,floatfix]{revtex4}
\usepackage{epsfig} \usepackage{graphics} \usepackage{bm}
\usepackage{amssymb}
\usepackage{graphicx}
\begin{document}

\preprint{Lebed-100}

\title{Inequivalence Between Passive Gravitational Mass and Energy
for a Quantum Body: Theory and Suggested Experiment}

\author{Andrei G. Lebed$^*$}

\affiliation{Department of Physics, University of Arizona, 1118 E.
4-th Street, Tucson, Arizona 85721, USA; \
lebed@physics.arizona.edu}

\begin{abstract}
Recently, we have suggested some semi-quantitative Hamiltonian for
an electron in a hydrogen atom in a weak gravitational field,
which takes into account quantum effects of electron motion in the
atom. We have shown that this Hamiltonian predicts breakdown of
the equivalence between passive electron gravitational mass and
its energy. Moreover, as has been shown by us, the latter
phenomenon can be experimentally observed as unusual emission of
radiation from an ensemble of the atoms, provided that they are
moved in the Earth's gravitational field with constant velocity by
some spacecraft. In this article, we derive the above-mentioned
Hamiltonian from the Dirac equation in a curved spacetime. It is
shown that it exactly coincides with the semi-quantitative
Hamiltonian, used in our previous papers. We extend the obtained
Hamiltonian to the case of a spacecraft, containing a macroscopic
ensemble of the atoms and moving with a constant velocity in the
Earth's gravitational field. On this basis, we discuss some
idealized and realistic experiments on the Earth's orbit. If such
(realistic) experiment is done, it will be the first direct
observation of quantum effects in the General Relativity.
\end{abstract}

\pacs{04.60.-m, 04.80.Cc}

\maketitle

%\begin{document}

\section{1. Introduction}

It is well known that creation of the Quantum Gravitation theory
is the most important step in developing of the so-called Theory
of Everything. This problem appears to be extremely difficult.
This is partially due to the fact that the foundations of quantum
mechanics and the General Relativity are very different and
partially due to the absence of the corresponding experimental
data. So far, quantum effects have been directly observed only in
the Newtonian variant of gravitation \cite{Collela,Nesvizh}. On
the other hand, such important quantum phenomenon in the General
Relativity as the Hawking radiation \cite{Hawking} is still very
far from its direct experimental discovery. In this complicated
situation, we have suggested two novel phenomena
\cite{Lebed-1,Lebed-2,Lebed-3,Lebed-4,Lebed-5}, which show that
passive and active gravitational masses of a composite quantum
body are not equivalent to its energy due to some quantum effects.
Moreover, we have also proposed experimental ways
\cite{Lebed-1,Lebed-2,Lebed-3,Lebed-4,Lebed-5} how to observe the
above mentioned phenomena. If one of these experiments is done, it
will be the first direct observation of quantum effects in the
General Relativity. In this paper, we concentrate our
consideration on passive gravitational mass of the simplest
composite quantum body - a hydrogen atom. The main our results are
derivations of two Hamiltonians from the Dirac equation in curved
spacetime of the General Relativity, which were semi-qualitatively
introduced by us in
Refs.\cite{Lebed-1,Lebed-2,Lebed-3,Lebed-4,Lebed-5}. Note that a
notion of passive gravitational mass of a composite body is not
trivial even in classical physics. As mentioned by Nordtvedt
\cite{Nordtvedt} and Carlip \cite{Carlip}, an external weak
gravitational field is coupled with the combination $m_e + m_p +
(3K + 2P)/c^2$, where c is the velocity of light, $m_e$ and $m_p$
are the bare masses of electron and proton, $K$ and $P$ are
kinetic and potential energies of an electron. Nevertheless, due
to the classical virial theorem averaged over time combination
$<2K + P>_t =0$ and, therefore, averaged over time electron
gravitational mass, $<m_g>_t$, satisfy the famous Einstein's
equation,
\begin{equation}
<m_g>_t = m_e + \biggl< \frac{3K + 2P}{c^2} \biggl>_t = m_e +
\biggl< \frac{K + P}{c^2} \biggl>_t = \frac{E}{c^2},
\end{equation}
where $E$ is the total electron energy. On this basis, in
Refs.\cite{Nordtvedt,Carlip}, the conclusion about the equivalence
between averaged over time passive gravitational mass and energy
was made. Recently, in
Refs.\cite{Lebed-1,Lebed-2,Lebed-3,Lebed-4,Lebed-5}, we have
considered a quantum hydrogen atom in a weak gravitational field.
We have suggested semi-quantitative derivation of its
post-Newtonian Hamiltonian, ignoring the so-called tidal
corrections, magnetic force, and all spin related phenomena. As a
result, we have made the following three statements about passive
gravitational mass \cite{Lebed-1,Lebed-2,Lebed-3,Lebed-4,Lebed-5}:
(a) The expectation value of the mass for a stationary electron
quantum state is equivalent to its energy; (b) For an individual
measurement, there exists small probability that the mass is not
equivalent to the energy; (c) The above mentioned inequivalence of
the mass and energy can be experimentally observed as a detection
of unusual radiation, emitted by a macroscopic ensemble of the
atoms, moved with constant velocity by spacecraft in the Earth's
gravitational field.

\section{2. Some of Our Previous Results
\cite{Lebed-1,Lebed-2,Lebed-3,Lebed-4,Lebed-5}} In Subsection 2.1,
we describe in brief semi-quantitative derivation of the quantum
Hamiltonian for a hydrogen atom in a weak gravitational field. In
Subsection 2.2, we discuss how the quantum virial theorem
\cite{Park} results in a surviving of the equivalence between the
expectation value of electron passive gravitational mass and its
energy. In Subsection 2.3, we show that the equivalence between
electron passive gravitational mass and its energy is broken with
a small probability during individual measurements of the mass.

\subsection{2.1. Semi-Quantitative Derivation of the Hamiltonian
\cite{Lebed-1,Lebed-2,Lebed-3,Lebed-4,Lebed-5}}

Let us write a standard interval, describing spacetime in a weak
field approximation outside some massive body \cite{Misner},
\begin{eqnarray}
ds^2 = - \biggl( 1 + 2\frac{\phi}{c^2} \biggl)(c dt)^2 + \biggl( 1
- 2 \frac{\phi}{c^2} \biggl) (dx^2 +dy^2+dz^2 ), \ \ \phi = -
\frac{GM}{R},
\end{eqnarray}
where $G$ is the gravitational constant, $M$ is the body mass, and
$R$ is the distance between its center and center of mass of a
hydrogen atom. Then, according to the General Relativity we can
introduce the following local proper spacetime coordinates,
\begin{equation}
x'=\biggl(1-\frac{\phi}{c^2} \biggl) x, \ \ \ y'=
\biggl(1-\frac{\phi}{c^2} \biggl) y, \ \ \
z'=\biggl(1-\frac{\phi}{c^2} \biggl) z , \ \ \ t'=
\biggl(1+\frac{\phi}{c^2} \biggl) t,
\end{equation}
where the Schr\"{o}dinger equation for electron in a hydrogen atom
can be approximately expressed in its standard form,
\begin{equation}
i \hbar \frac{\partial \Psi({\bf r'},t')}{\partial t'} = \hat H_0
(\hat {\bf p'},{\bf r'}) \ \Psi({\bf r'},t')  ,
\end{equation}
where
\begin{equation}
\hat H_0 (\hat {\bf p'},{\bf r'})= m_ec^2 - \frac{\hat {\bf
p'}^2}{2 m_e} - \frac{e^2}{r'} .
\end{equation}
We point out that Eqs.(2)-(5) are written in the so-called $1/c^2$
approximation. As to gravitation, we take into account only terms
of the order of $\phi/c^2$, which can be estimated as $10^{-9}$
near the Earth, and disregard terms of the order of $(\phi/c^2)^2
\sim 10^{-18}$. We also stress that, in Eqs.(4),(5), we do not
take into account the so-called tidal effects (i.e., we do not
differentiate gravitational potential $\phi$ with respect to
electron coordinates, ${\bf r}$ and ${\bf r'}$). Note that ${\bf
r}$ and ${\bf r'}$ correspond to electron positions in the center
of mass coordinate system, which we relate to proton position. In
Section 3, we show that, in fact, this means that we consider a
hydrogen atom as a point-like body and disregard all tidal terms
in the electron Hamiltonian, which are usually very small and are
of the order of $(r_B/R_0)|\phi/c^2| \sim 10^{-17}|\phi/c^2| \sim
10^{-26}$ in the Earth's gravitational field. Here, $r_B$ is the
so-called Bohr's radius and $R_0$ is the Earth's radius. In
Eqs.(4),(5), we also disregard magnetic force and all spin related
effects. Another our suggestion is that proton mass is very high,
$m_p \gg m_e$, and, thus, proton can be considered as a classical
particle, whose position is fixed and kinetic energy is
negligible. In this article, as usual, we consider the weak
gravitational field (2) as a perturbation in some inertial
coordinate system. The inertial coordinate system corresponds to
Minkowskii spacetime coordinates $(x,y,z,t)$ in Eq.(3), where it
is possible to obtain the following electron Hamiltonian from
Eqs.(3)-(5):
\begin{equation}
\hat H(\hat {\bf p},{\bf r}) = m_e c^2 + \frac{\hat {\bf
p}^2}{2m_e}-\frac{e^2}{r} + m_e  \phi + \biggl( 3 \frac{\hat {\bf
p}^2}{2 m_e} -2\frac{e^2}{r} \biggl) \frac{\phi}{c^2}.
\end{equation}
Note that the Hamiltonian (6) can be rewritten in more convenient
form,
\begin{equation}
\hat H(\hat {\bf p},{\bf r}) = m_e c^2 + \frac{\hat {\bf
p}^2}{2m_e} -\frac{e^2}{r} + \hat m_g (\hat {\bf p},{\bf r}) \phi
\ ,
\end{equation}
where we introduce the following electron passive gravitational
mass operator:
\begin{equation}
\hat m_g (\hat {\bf p},{\bf r})  = m_e  + \biggl(\frac{\hat {\bf
p}^2}{2m_e} -\frac{e^2}{r}\biggl) \frac{1}{c^2} + \biggl(2
\frac{\hat {\bf p}^2}{2m_e}-\frac{e^2}{r} \biggl) \frac{1}{c^2} ,
\end{equation}
which is proportional to electron weight operator in the weak
gravitational field (2). It is important that, in Eq.(8), the
first term corresponds to the bare electron mass, $m_e$, the
second term corresponds to the expected electron energy
contribution to the mass operator, whereas the third term is the
non-trivial virial contribution to passive gravitational mass
operator.

\subsection{2.2. Equivalence Between the Expectation Value of Mass and Energy
\cite{Lebed-1,Lebed-2,Lebed-3,Lebed-4,Lebed-5}}

In this Subsection, we discuss one important consequence of
Eqs.(7),(8). We stress that the operator (8) does not commute with
taken in the absence of the gravitational field electron energy
operator. Therefore, from the beginning, it seems that there is no
any equivalence between electron passive gravitational mass and
its energy. To establish their equivalence at a macroscopic level,
we consider a macroscopic ensemble of hydrogen atoms with each of
them being in a stationary ground state with energy $E_1$,
\begin{equation}
\Psi_1(r,t) = \Psi_1(r) \exp \biggl( \frac{-im_ec^2t}{\hbar}
\biggl) \exp \biggl( \frac{- iE_1t}{\hbar} \biggl)\ ,
\end{equation}
where $\Psi_1(r)$ is a normalized electron ground state wave
function in a hydrogen atom. As follows from Eq.(8), in this case,
the expectation value of passive gravitational mass operator per
one electron is
\begin{equation}
<\hat m_g > = m_e + \frac{ E_1}{c^2}  + \biggl< 2 \frac{\hat {\bf
p}^2}{2m_e}-\frac{e^2}{r} \biggl> \frac{1}{c^2} = m_e +
\frac{E_1}{c^2} ,
\end{equation}
where the expectation value of the virial term in Eq.(10) is zero
due to the quantum virial theorem \cite{Park}. Note that the
result (10) can be easily extended to any stationary quantum state
in a hydrogen atom. Thus, we can conclude that the equivalence
between passive gravitational mass and energy survives at a
macroscopic level for stationary quantum states.

\subsection{2.3. Breakdown of the Equivalence Between Passive Gravitational
Mass and Energy at a Microscopic Level
\cite{Lebed-1,Lebed-2,Lebed-3,Lebed-4,Lebed-5}}

Here, we describe a thought experiment, which directly
demonstrates inequivalence between passive gravitational mass and
energy at a microscopic level. Let us consider electron with a
ground state wave function (9) in a hydrogen atom, corresponding
to the absence of the gravitational field (2),
\begin{equation}
\hat H_0({\bf p},r) \Psi_1(r) = E_1 \Psi_1(r) , \ \ \ \hat
H_0({\bf p},r)= m_ec^2 + \frac{\hat {\bf p}^2}{2 m_e}-
\frac{e^2}{r}.
\end{equation}
Now, we take into account the gravitational field (2), as a
perturbation to the Hamiltonian (11),
\begin{equation}
\hat H(\hat {\bf p},{\bf r}) = \hat H_0({\bf p},r) + \hat m_g
(\hat {\bf p},{\bf r}) \phi \ ,
\end{equation}
where electron passive gravitational mass operator is given by
Eq.(8). Let us apply to the Hamiltonian (12) and its ground state
wave function, $\tilde \Psi_1(r)$, where
\begin{equation}
\hat H(\hat {\bf p},{\bf r}) \tilde \Psi_1(r) = \tilde E_1 \tilde
\Psi_1(r),
\end{equation}
\begin{equation}
\tilde \Psi_1(r) = \sum_n a_n \Psi_n(r) ,
\end{equation}
the standard quantum mechanical perturbation theory. Note that, in
Eq.(14), $\Psi_n (r)$ are normalized electron wave functions in a
hydrogen atom in the absence of the gravitation (2), corresponding
only to atomic levels $nS$ with energy $E_n$, due to a special
selection rule of the perturbation (8). It is possible to show
that, in accordance with the perturbation theory, coefficient
$a_1$ and correction to energy of the ground state can be
expressed as:
\begin{equation}
a_1 \simeq 1, \ \ \ \tilde E_1 = \biggl[1 + \frac{\phi(R)}{c^2}
\biggl] E_1,
\end{equation}
where the last term in Eq.(15) corresponds to the famous red shift
in a gravitational field. It manifests the expected contribution
to passive gravitational mass due to electron binding energy in a
hydrogen atom. Note that during derivation of Eq.(15), we have
used the quantum virial theorem \cite{Park}, which states that
\begin{equation}
\int \Psi^*_1(r) \biggl( 2 \frac{\hat {\bf p}^2}{2 m_e} -
\frac{e^2}{r} \biggl) \Psi_1(r) d^3 {\bf r} = 0 .
\end{equation}
As to the coefficients $a_n$ with $n \neq 1$ in Eq.(14), they can
be expressed as functions of the virial operator matrix elements,
\begin{equation}
a_n = \biggl[ \frac{\phi(R)}{c^2} \biggl] \biggl(
\frac{V_{n,1}}{E_1-E_2} \biggl), \ \ \ V_{n,1} = \int \Psi^*_n(r)
\biggl( 2 \frac{\hat {\bf p}^2}{2 m_e} - \frac{e^2}{r} \biggl)
\Psi_1(r) d^3 {\bf r} .
\end{equation}
Here, we point out that the wave function (14)-(17), corresponding
to electron ground state in the presence of the gravitational
field (2), is a series of eigenfunctions of electron energy
operator in the absence of the field. Therefore, if we measure
energy in electron quantum state (14)-(17) by means of operator
(11), we obtain the following quantized values:
\begin{equation}
E(n) = m_e c^2 + E_n,
\end{equation}
where we disregard the red shift effect. Other words, we can
conclude that with probability close to 1 [see Eq.(15)], the
Einstein equation, $E = m_e c^2 + E_1$, survives in our case,
whereas with small probabilities,
\begin{equation}
P_n = |a_n|^2 = \biggl[\frac{\phi(R)}{c^2}\biggl]^2
\frac{V^2_{n,1}}{(E_n -E_1)^2} , \ \ \ n \neq 1 ,
\end{equation}
it is broken. The reason for this breakdown is that, as follows
from Eqs.(14)-(17), electron wave function with definite passive
gravitational mass is not characterized by definite energy in the
absence of the gravitational field. Below, we would like to
discuss two points, related to the fact that the probabilities
(19) are of the second order magnitude with respect to small
parameter $(\phi/c^2)$. First point - the accuracy of our
calculations. According to quantum mechanics, it is enough to
calculate wave function (14) in a linear approximation with
respect to small parameter $\phi/c^2$ to obtain probabilities
(19). Second point is related to physical meaning of our results.
In this article, we do not show that the average passive
gravitational mass of electron is not equal to its energy, since
we do not take into account terms of the order of $(\phi/c^2)^2$
in Eqs.(2),(3) as well as we do not calculate correction of the
order of $(\phi/c^2)^2$ to the ground state of a hydrogen atom.
Here, it is important that quantum mechanics, as known, is not a
science about average values, but it is a science about
probabilities of individual measurements. What we really show is
that electron with definite gravitational mass (8) can be excited
with small probabilities (19) to high energy levels (18) during
quantum measurement of its energy by means of operator (11). Why
is that so important? The answer is the following: high energy
levels are quasi-stationary and we can expect photons emission
from a macroscopic ensemble of the atoms. For description of the
corresponding experiment, see Section 4.

\section{3. Results}

In this Section, we present in detail the main results of the
current article - careful derivations of two Hamiltonians from the
Dirac equation in a curved spacetime of the General Relativity. In
Subsection 3.1, we present and comment the Hamiltonian for a
hydrogen atom, derived in Ref.\cite{Fischbach}, where the
so-called tidal terms are taken into account as well as motion of
a center of mass of the atom. In Subsection 3.2, we derive from
the Hamiltonian \cite{Fischbach} the considered in the previous
Section Hamiltonian (7),(8), whereas, in Subsection 3.3, we derive
the Hamiltonian, which is used later for description of the
suggested experiment (see Section 4).

\subsection{3.1. The Most General Hamiltonian \cite{Fischbach}}

In Ref.\cite{Fischbach}, the mixing effect between even and odd
wave functions in a hydrogen atom (i.e., the so-called
relativistic Stark effect) was studied in an external
gravitational field. A possibility of a center of mass motion of
the atom was taken into account and the corresponding Hamiltonian
was derived in $1/c^2$ approximation. The peculiarity of the
calculations in the above mention article was that not only terms
of the order of $\phi/c^2$ were calculated, as in our case, but
also terms of the order of $\phi'/c^2$, where $\phi'$ is a
symbolic derivative of $\phi$ with respect to reciprocal electron
coordinates in the atom. According to the existing tradition, we
call the latter terms tidal ones. The Hamiltonian (3.24), obtained
in Ref. \cite{Fischbach} for the corresponding Schr\"{o}dinger
equation, can be written as a sum of the following four terms:
\begin{equation}
\hat H (\hat {\bf P}, \hat {\bf p}, \tilde {\bf R},r)= \hat H_0
(\hat {\bf P}, \hat {\bf p}, r) + \hat H_1 (\hat {\bf P}, \hat
{\bf p}, \tilde {\bf R},r) + \hat H_2 (\hat {\bf p}, {\bf r}) +
\hat H_3 (\hat {\bf P}, \hat {\bf p},\tilde {\bf R},r) ,
\end{equation}
where
\begin{equation}
\hat H_0 (\hat {\bf P}, \hat {\bf p}, r) = m_e c^2 + m_p c^2 +
\biggl[\frac{\hat {\bf P}^2}{2(m_e + m_p)} + \frac{\hat {\bf
p}^2}{2 \mu} \biggl] - \frac{e^2}{r} ,
\end{equation}

\begin{equation}
\hat H_1 (\hat {\bf P}, \hat {\bf p}, \tilde {\bf R}, r) =
\biggl\{ m_e c^2 + m_p c^2 +  \biggl[3 \frac{\hat {\bf P}^2}{2(m_e
+ m_p)} + 3 \frac{\hat {\bf p}^2}{2 \mu} - 2 \frac{e^2}{r}
\biggl]\biggl\} \biggl( \frac{\phi  - {\bf g}\tilde {\bf R}}{c^2}
\biggl),
\end{equation}

\begin{equation}
\hat H_2 (\hat {\bf p}, {\bf r}) = \frac{1}{c^2}
\biggl(\frac{1}{m_e}-\frac{1}{m_p} \biggl)[-({\bf g}{\bf r}) \hat
{\bf p}^2 + i \hbar {\bf g} \hat {\bf p}] + \frac{1}{c^2} {\bf g}
\biggl(\frac{\hat {\bf s_e}}{m_e} - \frac{\hat {\bf s_p}}{m_p}
\biggl) \times \hat {\bf p} + \frac{e^2 (m_p-m_e)}{2(m_e+m_p)c^2}
\frac{{\bf g}{\bf r}}{r},
\end{equation}

\begin{equation}
\hat H_3 (\hat {\bf P}, \hat {\bf p}, \tilde {\bf R}, r) =
\frac{3}{2}\frac{i \hbar {\bf g}{\bf P}}{(m_e+m_p)c^2}
+\frac{3}{2} \frac{{\bf g}{\bf(s_e+s_p)}\times {\bf
P}}{(m_e+m_p)c^2} - \frac{({\bf g}{\bf r})({\bf P}{\bf p})+({\bf
P}{\bf r})({\bf g}{\bf p})-i\hbar {\bf g}{\bf P}}{(m_e+m_p)c^2},
\end{equation}
where ${\bf g}=-G \frac{M}{R^3} {\bf R}$. In Eqs.(20)-(24),
$\tilde {\bf R}$ and ${\bf P}$ stand for coordinate and momentum
of a hydrogen atom center of mass, respectively. Whereas, ${\bf
r}$ and ${\bf p}$ stand for reciprocal electron coordinate and
momentum in the center of mass coordinate system; $\mu = m_e m_p
/(m_e + m_p)$ is the so-called reduced electron mass. Note that
$\hat H_0 (\hat {\bf P}, \hat {\bf p}, r)$ represents the
Hamiltonian of a hydrogen atom in the absence of a gravitational
field. On the other hand, $\hat H_1 (\hat {\bf P}, \hat {\bf p},
\tilde {\bf R}, r)$ describes couplings of the bare electron and
proton masses to the gravitational field (2) as well as couplings
of kinetic and potential energies to the field. The Hamiltonians
$\hat H_2 (\hat {\bf p}, {\bf r})$ and $\hat H_3 (\hat {\bf P},
\hat {\bf p}, \tilde {\bf R}, r)$ contain tidal effects.

\subsection{3.2. Derivation of the Hamiltonian (7),(8)}

Let us strictly derive the Hamiltonian (7),(8), which has been
semi-quantitatively derived in Section 2, from the more general
Hamiltonian (20)-(24). First of all, for simplicity we use the
approximation, where $m_p \gg m_e$, and, thus, $\mu = m_e$ , which
allow to consider proton as a heavy classical particle. In
addition, in this Subsection, we derive the Hamiltonian of a
hydrogen atom, whose center of mass is at rest in the
gravitational field (2). Therefore, we can disregard center of
mass momentum and center of mass kinetic energy. As a result, in
the case under consideration, the first two contributions to the
total Hamiltonian (20)-(24) can be rewritten in the following way:
\begin{equation}
\hat H_0 (\hat {\bf p}, r) = m_e c^2 + \frac{\hat {\bf p}^2}{2m_e}
- \frac{e^2}{r}
\end{equation}
and
\begin{equation}
\hat H_1 (\hat {\bf p}, r) =  \biggl\{ m_e c^2 + \biggl[3
\frac{\hat {\bf p}^2}{2 m_e} - 2 \frac{e^2}{r} \biggl]\biggl\}
\biggl( \frac{\phi}{c^2} \biggl),
\end{equation}
where we place the center of mass of the atom at point $\tilde
{\bf R} = 0$. Let us consider the first tidal contribution (23) to
the total Hamiltonian (20). Note that $|{\bf g}| \simeq
|\phi|/R_0$, where $R_0$ is the Earth's radius. In addition, in a
hydrogen atom, $|{\bf r}| \sim \hbar / |{\bf p}| \sim r_B$ and
${\bf p}^2/(2m_e) \sim e^2/r_B$. These values allow us to estimate
the Hamiltonian (23) as $H_2 \sim (r_B/R_0) (\phi/c^2) (e^2/r_B)
\sim 10^{-17} (\phi/c^2) (e^2/r_B)$, which is $10^{-17}$ smaller
than $H_1 \sim (\phi/c^2) (e^2/r_B)$ and $10^{-8}$ smaller than
the second correction with respect to the small parameter
$|\phi|/c^2$, which is omitted. Thus, we can disregard the
contribution (23) to the total Hamiltonian (20)-(24). Moreover, we
pay attention that the second tidal contribution (24) to the total
Hamiltonian is exactly zero in the case, where ${\bf P}=0$,
considered in this Subsection. At this point, we can conclude that
the Hamiltonian (25),(26), derived above, exactly coincides with
that, semi-quantitatively derived by us in
Refs.\cite{Lebed-1,Lebed-2,Lebed-3,Lebed-4,Lebed-5} [see
Eqs.(7),(8)].

\subsection{3.3. Derivation of the Hamiltonian to Describe the Suggested
Experiment \cite{Lebed-1,Lebed-2,Lebed-3,Lebed-4,Lebed-5}}

In this Subsection, we derive the Hamiltonian for the case, where
proton is fixed in a inertial coordinate system, moving from the
Earth with a constant velocity $u \ll c$. We relate such inertial
system to a spacecraft in accordance with the experiment,
suggested in Refs. \cite{Lebed-1,Lebed-2,Lebed-3,Lebed-4,Lebed-5}
and described in the next Section. As in the previous Subsection,
we use here the relationship $m_p \gg m_e$, which allows us to
consider proton as a classical particle as well as to disregard
its kinetic energy. In the case under consideration, the first two
contributions to the total Hamiltonian (20) in the inertial
system, related to the spacecraft, can be written as
\begin{equation}
\hat H_0 (\hat {\bf p}, r) = m_e c^2 + \frac{\hat {\bf p}^2}{2m_e}
- \frac{e^2}{r}
\end{equation}
and
\begin{equation}
\hat H_1 (\hat {\bf p}, r) =  \biggl\{ m_e c^2 + \biggl[3
\frac{\hat {\bf p}^2}{2 m_e} - 2 \frac{e^2}{r} \biggl]\biggl\}
\biggl[ \frac{\phi(R_0 + ut)}{c^2} \biggl].
\end{equation}
Note that the contribution (23) to the total Hamiltonian (20)-(24)
is estimated in the same way, as in the previous Subsection, and,
thus, can be disregarded. As to the contribution (24), it is
non-zero in our case and has to be estimated separately. We
consider a realistic situation, where our spacecraft moves with
constant velocity $u \ll \alpha c$, where $\alpha$ is the fine
structure constant and, thus, $\alpha c$ is a typical value of
electron velocity in the Bohr's model of a hydrogen atom. In this
case, the value of the Hamiltonian (24) can be estimated as
$(r_B/R_0) (m_p/m_e) (\phi/c^2) \sim 10^{-14} (\phi/c^2)$ and,
therefore, can be disregarded. It is important that our suggestion
for the experiment is to observe photons emitted by the atoms,
excited in the gravitational field, therefore, it is enough to
keep in the Hamiltonian (27),(28) only the virial term, which
gives non-zero transitions between different energy levels:
\begin{equation}
\hat H_0 (\hat {\bf p}, r) = m_e c^2 + \frac{\hat {\bf p}^2}{2m_e}
- \frac{e^2}{r}
\end{equation}
and
\begin{equation}
\hat H_1 (\hat {\bf p}, r) =  \biggl(2 \frac{\hat {\bf p}^2}{2
m_e} -  \frac{e^2}{r} \biggl) \biggl[ \frac{\phi(R_0 + ut)}{c^2}
\biggl].
\end{equation}
Here, we pay attention that the Hamiltonian (29),(30) exactly
coincides with that, semi-quantitatively introduced in
Refs.\cite{Lebed-1,Lebed-2,Lebed-3} and derived from the
Lagrangian \cite{Nordtvedt} in Refs.\cite{Lebed-3,Lebed-5}.

\section{4. Suggested Idealized Experiment \cite{Lebed-1,Lebed-2,Lebed-3,Lebed-4,Lebed-5}}

Below, we discuss an idealized experiment to observe the breakdown
of the equivalence between passive gravitational mass and energy,
which is a consequence of the derived Hamiltonians (29),(30). We
assume that we have a macroscopic ensemble of hydrogen atoms with
each of them being in a ground state at initial moment of time,
$t=0$. All these atoms are supposed to be placed in a small
spacecraft and all protons are supposed to have fixed positions in
the spacecraft. Then, we suggest that the spacecraft is located at
distance $R^*$ from the Earth's center and is at rest for some
time at $t<0$ to prepare the following equilibrium ground state
electron wave function in the atoms \cite{Lebed-5}:
\begin{equation}
\tilde{\Psi}_1(r,t) = \biggl( 1- \frac{\phi^*}{c^2} \biggl)^{3/2}
\Psi_1 \biggl[ \biggl( 1- \frac{\phi^*}{c^2} \biggl)r \biggl] \exp
\biggl[ -im_ec^2 \biggl( 1+ \frac{\phi^*}{c^2} \biggl) \frac{t}
{\hbar} \biggl] \ \exp \biggl[ -iE_1 \biggl( 1+ \frac{\phi^*}{c^2}
\biggl)\frac{t}{\hbar} \biggl] \ ,
\end{equation}
where $\phi^*=\phi(R^*)$. It is important that, in the idealized
experiment, we consider non-interacting hydrogen atoms, therefore,
we can first study behavior of a single atom. We recall that
proton position is fixed in the aircraft, which first accelerates
for a short time and then moves long enough time with constant
velocity $u \ll \alpha c$, which is directed from the center of
the Earth. The derived above Hamiltonian (29),(30) does not allow
us to consider a short period of the spacecraft acceleration and
we disregard it. Instead, we concentrate on a consideration of the
main part of the spacecraft trajectory, where it moves with
constant velocity, which can be studied by means of the
Hamiltonian (29),(30). Note that, on the latter part of the
spacecraft trajectory, electrons are excited by gravitational
potential due to its change in time in Eq.(30). Therefore, on the
main part of the trajectory, we can expect the following electron
wave function \cite{Lebed-5}:
 \begin{equation}
\tilde{\Psi}(r,t) = \biggl( 1- \frac{\phi^*}{c^2} \biggl)^{3/2}
\sum^{\infty}_{n=1} \tilde{a}_n(t) \Psi_n \biggl[ \biggl( 1-
\frac{\phi^*}{c^2} \biggl)r \biggl] \exp \biggl[ -im_ec^2 \biggl(
1+ \frac{\phi^*}{c^2} \biggl) \frac{t}{\hbar} \biggl] \exp \biggl[
-iE_n \biggl( 1+ \frac{\phi^*}{c^2} \biggl) \frac{t}{\hbar}
\biggl] ,
\end{equation}
where the perturbation (30) for a hydrogen atom can be expressed
as
\begin{equation}
\hat U ({\bf r},t) =\frac{\phi(R^*+ut)-\phi(R^*)}{c^2} \ \hat V
(\hat {\bf p}, r) , \ \ \ \hat V (\hat {\bf p}, r) = \biggl(2
\frac{\hat {\bf p}^2}{2m_e} - \frac{e^2}{r} \biggl) .
\end{equation}
We point out that in the spacecraft, which moves with constant
velocity, each electron experiences gravitational force action,
${\bf F} = m_e {\bf g}$. It is important that, as possible to
show, such force just slightly changes electron energy levels in a
hydrogen atom and does not produce electron excitations. Here, we
apply the standard time-dependent quantum mechanical perturbation
theory to calculate electron amplitudes $\tilde a_n(t)$ in Eq.(32)
for $n \neq 1$,
\begin{equation}
\tilde{a}_n(t)= -\frac{V_{1,n}}{\hbar \omega_{n,1}c^2}
\biggl\{[\phi(R^*+ut)-\phi(R^*)]  \exp(i \omega_{n,1}t) -
\frac{u}{i \omega_{n,1}} \int^t_0 \frac{d \phi(R^*+ut)}{dR^*}
d[\exp(i \omega_{n,1}t)]\biggl\} ,
\end{equation}
where
\begin{equation}
\omega_{n,1} = \frac{E_n - E_1}{\hbar}
\end{equation}
and $V_{1,n}$ is given by Eq.(17). We pay attention that, under
conditions of the suggested experiment, the following inequality
is obviously valid:
\begin{equation}
u \ll \omega_{n,1} R \sim \alpha c \frac{R}{r_B} \sim 10^{13} c.
\end{equation}
It is easy to see that Eq.(36) means that the second term in
Eq.(34) is much less than the first one and, thus, can be omitted
below:
\begin{equation}
\tilde{a}_n(t)= -\frac{V_{1,n}}{\hbar \omega_{n,1}c^2}
[\phi(R^*+ut)-\phi(R^*)]  \exp(i \omega_{n,1}t) \ , \ \ \ n \neq
1.
\end{equation}
Note that, as directly follows from Eq.(37), there are exist
non-zero probabilities to find electron on the levels nS with $n
\neq 1$ in Eq.(18) due to electron excitations during the
experiment. If the excited levels of the atom were strictly
stationary, then the corresponding probabilities would be:
\begin{equation}
\tilde{P}_n(t)= |\tilde a_n(t)| = \biggl( \frac{V_{1,n}}{\hbar
\omega_{n,1}} \biggl)^2 \frac{[\phi(R^*+ut)-\phi(R^*)]^2}{c^4} =
\biggl( \frac{V_{1,n}}{\hbar \omega_{n,1}} \biggl)^2
 \frac{[\phi(R')-\phi(R^*)]^2}{c^4}, \ \ \  n \neq 1,
\end{equation}
where $R'=R^*+ut$. As well known, in fact, the excited levels of
any atom are quasi-stationary ones and, thus, they spontaneously
decay with time. In this situation, it is necessary to measure the
spontaneous emission of radiation, corresponding to $n \neq 1$ in
Eq.(18), which directly manifests the breakdown of the Einstein's
equation for passive gravitational mass and energy. [Here, we
point out that the dipole matrix elements for optical transitions
$nS \rightarrow 1S$ are zero. Under such conditions, we expect
dipole transitions, such as $3S \rightarrow 2P$ and $2P
\rightarrow 1S$, as well as quadrupole transitions, such as $2S
\rightarrow 1S$, etc. Let us estimate the probabilities (38). It
is easy to see that, by using spacecraft, we can reach the
condition $|\phi(R')| \ll |\phi(R^*)|$. In this case Eq.(38)
coincides with Eq.(19), which reflects their common physical
meaning, and can be written as
\begin{equation}
\tilde{P}_n = \biggl( \frac{V_{1,n}}{E_n - E_1} \biggl)^2
\frac{\phi^2(R^*)}{c^4}  \simeq  0.49 \times 10^{-18} \biggl(
\frac{V_{n,1}}{E_n-E_1} \biggl)^2 .
\end{equation}
Note that, in Eq.(39), we use the following values of the Earth's
parameters: $R_0 \simeq 6.36 \times 10^6 m$ and $M \simeq 6 \times
10^{24} kg$. It is important that probabilities (39) are of the
second order of magnitude with respect to the small parameter,
$|\phi/c^2| \ll 1$, and, thus, small, $P_n \sim 10^{-18}$. On the
hand, it is not too small and the corresponding number of the
excited electrons in a microscopic ensemble of the atoms can be
very large. Indeed, $V_{n,1}/(E_n-E_1) \sim 1$ and the Avogadro
number is $N_A = 6 \times 10^{23}$. Therefore, the number of
excited electrons for 1000 moles of the atoms is estimated as
\begin{equation}
N_{n,1} = 2.95 \times 10^{8} \biggl( \frac{V_{n,1}}{E_n-E_1}
\biggl)^2 , \ \ \ N_{2,1} = 0.9 \times 10^8,
\end{equation}
where $N_{n,1}$ determines the number of electrons excited from
energy level $1S$ to energy level $nS$. We hope that the
corresponding photons, for such large amount of the excited atoms,
can be experimentally detected.

\section{5. Discussion}

In this Section, we discuss in brief how to carry out a real
experiment in space to detect those seldom events, where the
equivalence between passive gravitational mass and energy is
broken. We recall that all our calculations in this article have
been done for an atomic hydrogen in the approximation of
non-interacting atoms. In addition, we have suggested that all
protons are fixed in the spacecraft inertial coordinate system.
The best realization of these ideal conditions seems to be
fulfilled in solid Helium-4, where He atoms are connected by van
der Waals' forces. On the other hand, the suggested phenomenon is
very general and can be observed for any types of excitations: in
solids, in nuclei \cite{Crowell}, etc. For instance, we suggest to
measure conductivity in semiconductors to detect the excited
electrons.

\section{6. Conclusions}

To conclude, we point out that we have suggested to detect
photons, emitted by excited electrons in a macroscopic ensemble of
the atoms, provided that the atoms are moved in the Earth's
gravitational field with constant velocity, $u \ll \alpha c$, by a
spacecraft. Our proposal is very different from proposals,
discussed before, where small corrections to electron energy
levels are calculated for a hydrogen atom, supported in a
gravitational field \cite{Fischbach}, or for a free falling atom
\cite{Parker, Pinto}. Account of the above mentioned small
corrections to energy levels cannot change significantly the
number of electrons, excited due to spacecraft motion in an
gravitational field (2), calculated in the paper. It is also very
important that the suggested experiment is not a "free fall"
experiment. For free falling atoms, the effects, discussed by us
(in this paper and in Refs. \cite{Lebed-1, Lebed-2, Lebed-3,
Lebed-4, Lebed-5}), presumably disappear since free falling atoms
feel only the second derivative of a gravitational potential.
Therefore, we think that comparison of our effects with "free
fall" astronomical data, performed in Ref. \cite{Crowell}, is not
appropriate. In particular, in our opinion, the conclusion of Ref.
\cite{Crowell} that our theoretical results contradict to the
existing experimental astronomical data is incorrect.

%%%%%%%%%%%%%%%%%%%%%%%%%%%%%%%%%%%%%%%%%%
\vspace{6pt}

%%%%%%%%%%%%%%%%%%%%%%%%%%%%%%%%%%%%%%%%%%
\acknowledgments{We are thankful to N.N. Bagmet (Lebed), V.A.
Belinski, Steven Carlip, Fulvio Melia, Douglas Singleton, and V.E.
Zakharov for fruitful and useful discussions.}

\vspace{8pt}

$^*$Also at: L.D. Landau Institute for Theoretical Physics, RAS,
2 Kosygina Street, Moscow 117334, Russia.

\end{document}